\documentclass{article}

\usepackage{subfig}

\usepackage{arxiv}
\usepackage{amsmath}
\usepackage[utf8]{inputenc} 
\usepackage[T1]{fontenc}    
\usepackage{hyperref}       
\usepackage{url}            
\usepackage{booktabs}       
\usepackage{amsfonts}       
\usepackage{nicefrac}       
\usepackage{microtype}      
\usepackage{lipsum}		
\usepackage{graphicx}
\usepackage{natbib}
\usepackage{doi}

\title{Enhancing Electrocardiography Data Classification Confidence: A Robust Gaussian Process Approach (MuyGPs)}

\author{ \href{https://orcid.org/0000-0002-1537-0907}{\includegraphics[scale=0.06]{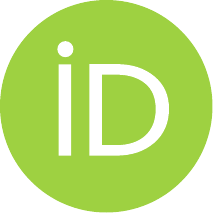}\hspace{1mm}Ukamaka V. Nnyaba \textsuperscript{\textdagger} }\thanks{Corresponding author} \\
	Department of Mathematics and Statistics\\
	Auburn University\\
	Auburn, AL, 36849, USA \\
	\texttt{uvn0001@auburn.edu} \\
	\And
	\href{https://orcid.org/0000-0001-8188-5477}{\includegraphics[scale=0.06]{orcid.pdf}\hspace{1mm}Hewan M. Shemtaga} \thanks{These authors contributed equally.}\\
	Department of Mathematics and Statistics\\
	Auburn University\\
	Auburn, AL, 36849 \\
	\texttt{hshemtaga@auburn.edu} \\
	\AND
	David W. Collins \\
	Department of Mathematics and Statistics\\
	Auburn University\\
	Auburn, AL, 36849\\
	\texttt{dcw0036@auburn.edu} \\
	\And
	Amanda L. Muyskens \\
	Computational Engineering Division \\
	Lawrence Livermore National Laboratory\\ 
    Livermore, CA 94550 \\
	\texttt{muyskens1@llnl.gov} \\
	\And
	Benjamin W. Priest \\
	Applied Scientific Computing \\
	Lawrence Livermore National Laboratory\\ 
    Livermore, CA 94550 \\
	\texttt{priest2@llnl.gov} \\
    \AND
	Nedret Billor \\
	Department of Mathematics and Statistics\\
	Auburn University\\
	Auburn, AL, 36849 \\
	\texttt{billone@auburn.edu} \\
}

\date{}

\renewcommand{\headeright}{}
\renewcommand{\undertitle}{}
\renewcommand{\shorttitle}{Enhancing Electrocardiography Data Classification Confidence}

\begin{document}
\maketitle

\begin{abstract}
Analyzing electrocardiography (ECG) data is essential for diagnosing and monitoring various heart diseases. The clinical adoption of automated methods requires accurate confidence measurements, which are largely absent from existing classification methods.
In this paper, we present a robust Gaussian Process classification hyperparameter training model (MuyGPs) for discerning normal heartbeat signals from the signals affected by different arrhythmias and myocardial infarction. We compare the performance of MuyGPs with traditional Gaussian process classifier as well as conventional machine learning models, such as, Random Forest, Extra Trees, k-Nearest Neighbors and Convolutional Neural Network. Comparing these models reveals MuyGPs as the most performant model for making confident predictions on individual patient ECGs. Furthermore, we explore the posterior distribution obtained from the Gaussian process to interpret the prediction and quantify uncertainty. In addition, we provide a guideline on obtaining the prediction confidence of the machine learning models and quantitatively compare the uncertainty measures of these models. Particularly, we identify a class of less-accurate (ambiguous) signals for further diagnosis by an expert.
\end{abstract}


\section{Introduction}\label{sec:introduction}
One of the major threats to human health is cardiovascular disease (CVD) \cite{cvd}. 
Over the past two decades, heart disease has accounted for 16\% of all deaths, making it a primary cause of death globally, according to the most recent World Health Statistics report \cite{whs}. Given the significant impact of heart disease on life expectancy, it is imperative to devise means for its detection and diagnosis. One of the diagnostic methods for detecting cardiovascular diseases is electrocardiography (ECG). ECG is an easy, cost-effective, non-invasive, highly efficient, and useful tool for monitoring and identifying rhythmic irregularities of the heart commonly known as arrhythmias, by measuring the electrical activity of the heart \cite{jay}. The ECG captures these electrical activities through a series of electrodes strategically placed on the skin, detecting and recording the electrical signals as they travel through the heart. The morphological changes of the ECG and the depolarization of the myocardium can be useful in the diagnosis of heart disease.
However, due to the complexity of ECG data, manual identification of these changes is very difficult, necessitating the broad medical expertise of physicians. Even then, physicians are not infallible. In the United States, most of the roughly 40 million ECGs taken annually are read by non-cardiologists \cite{noncardio}. These non-cardiologist physicians often have subpar accuracy and take significantly longer than their cardiologist counterparts \cite{competency}. Even when cardiologists do have the time to work on ECG tasks, they only correctly identify between 53\% and 96\% of ECG abnormalities across a variety of sources \cite{competency}. As the manual analysis by physicians is time-consuming, laborious and prone to error, it is desirable to design an ECG arrhythmia classification model that is fast, accurate, and confident. In light of this, many researchers have applied various algorithms to help detect these diseases, thereby reducing the workload of physicians and improving the efficiency of the diagnosis.
\vskip0.1truecm 
 Some of the classical and deep learning approaches for the ECG heartbeat classification are as follows: Naaz \textit{et al.} \cite{moha_dt} proposed an optimized decision tree and adaptive boosted optimized decision tree for classifying six types of heartbeats, achieving an accuracy of 98.77\% in ECG arrhythmia classification. To optimize the parameters of machine learning models for ECG classification, Hassaballah \textit{et al.} \cite{ecg_ml} integrated a recent metaheuristic optimization (MHO) algorithm, the Marine Predator Algorithm, proposed by Faramarzi \textit{et al.} \cite{marine}. This approach significantly improved the performance of the classifiers. Zhang \textit{et al.} \cite{ecg_cnn1} proposed a deep convolutional neural network (CNN) structure with $3\times 3$ convolutional kernel, achieving 98.8\%  accuracy on a subset of ECG arrhythmia classification. Pham \textit{et al.} \cite{ecg_cnn2} used residual blocks and an Evolving Normalization - Activation function (EVO) in their CNN model to achieve an accuracy of 98.56\% in the same classification task.

\vskip0.1truecm
These approaches and prior studies in ECG classification have often overlooked the evaluation and management of uncertainty associated with their predictions, focusing primarily on classification performance without considering the reliability and efficacy of the models. While research on uncertainty quantification remains limited, some recent studies have addressed this area. Zhang \textit{et al.} \cite{zhang22} estimated uncertainties in ECG classification using Monte Carlo dropout simulations. They also categorized predictions with uncertainty under a given threshold into a separate 'uncertain' category. Belen \textit{et al.} \cite{jbelen} used a variational encoder network to study uncertainty in classifying atrial fibrillation - a common type of arrhythmia, by conducting multiple passes of input through the network to build a distribution. Barandas \textit{et al.} \cite{ecg_uq} compared five different uncertainty quantification methods using the same deep neural network architecture in multi-label ECG classification.  Their study suggested ensemble-based methods resulted in more robust uncertainty estimation compared to single network or stochastic methods. Gawlikowski \textit{et al.} presented a comprehensive overview of uncertainty estimation in neural networks, focusing on deterministic neural networks, Bayesian neural networks (BNNs), ensembles of neural networks, and test-time data augmentation approaches
\cite{uq_survey}. 
\vskip0.1truecm 
While ensemble methods and test-time augmentation methods are relatively easy to apply, Bayesian model approaches such as Gaussian processes (GPs) deliver a clear description of the uncertainties on the model parameters and also provide a more profound theoretical basis. However, these approaches often face limitations in computational effort and memory consumption in real-life applications.
To address this, Muyskens \textit{et al.} \cite{muygps} developed a scalable approximate method (MuyGPs) for training stationary GP hyperparameters that give both accurate predictions and  efficiency, while maintaining accurate uncertainty quantification. This method has been successfully applied in classifying optical telescope images to distinguish stars from galaxies \cite{galaxy1, galaxy11} and identifying galaxy blends 
\cite{galaxy2}. As a result, the MuyGPs model compared favorably to other machine-learning models that were compared, while producing posterior distributions to quantify classification uncertainty. 

\vskip0.1truecm 
Motivated by the work of Muyskens \textit{et al.} \cite{muygps}, in this paper, we explore MuyGPs for discerning normal heartbeat signals from those affected by various arrhythmias and myocardial infarction. Given the demonstrated effectiveness of CNN models on ECG datasets and advantageous GP method, which  conveniently provides both predictions and  associated uncertainty, we integrate CNN with MuyGPs (CNN-MuyGPs). We show that MuyGPs and CNN-MuyGPs models compare favorably to other machine learning models such as extra trees, random forests, and k-nearest neighbor (kNNs) algorithms in terms of computational efficiency, accuracy, and scalability. Furthermore, we demonstrate how the posterior distribution obtained from GPs can be used to interpret predictions, quantify uncertainty, and provide a guideline for the prediction confidence of these machine learning algorithms. By using this uncertainty, we can identify a subset of less accurate (ambiguous) signals for further verification or diagnosis by a medical expert. This process further improves the performance of the classifiers and describes procedures to analyze and label large ECG signals reliably incorporating a level of automation.

\vskip0.1truecm 
Section \ref{sec2} is divided into two sections, 'Materials', which describes the ECG datasets used in this study and explores the dataset; and 'Methods', which provides a comprehensive overview of the Gaussian process models and the machine learning models employed in this study. Section \ref{sec4} demonstrates a comparative analysis of the performance of the models used in this study. Section \ref{sec3} outlines the procedures for uncertainty estimation procedures for predictions obtained from both the GP models and the machine learning models, and compares the performances of these models on uncertainty estimation. The findings from these analyses are then summarized in Section \ref{sec5}.

\section{Materials and Methods}\label{sec2}
\subsection{Materials}
\subsubsection{Dataset}
Our study uses datasets derived from the MIT-BIH Arrhythmia and the PTB Diagnostic ECG Databases \cite{mit_bih,ptb,physionet}. The MIT-BIH Arrhythmia Database contains ECG recordings from 47 subjects, which were collected by the BIH Arrhythmia Laboratory between 1975 and 1979. The recordings were digitized at 360 samples per second per channel, with an 11-bit resolution over a 10 mV range. Moreover, the annotations for each record were made independently by at least two cardiologists, with any disagreements being resolved. The PTB Diagnostic dataset contains 549 records from 290 subjects, with each signal digitized at 1000 samples per second and a 16 bit resolution over a 16.384 mV range. 

In the PTB Diagnostic dataset, the myocardial infarction and healthy control categories were used for binary classification. However, for MIT-BIH Arrhythmia dataset, the five beat annotation categories obtained in \cite{preprocess} were used for multi-class classification purpose. These categorizations were done in accordance with Association for the Advancement of Medical Instrumentation (AAMI) EC57 standard \cite{amdi}. For a summary of these categorizations, refer to Table ~\ref{tab2}. Moreover, in \cite{preprocess}, the authors developed an effective method for extracting respiratory rate (RR) intervals and made the preprocessed dataset publicly available in \cite{ecg_data}. In our study, we utilize these preprocessed PTB and MIT-BIH datasets for binary and multi-class classification. Table~\ref{tab1} summarizes the content of these datasets.

\begin{table}
\centering
\caption{Summary of mappings between beat annotations and AAMI EC57 \cite{amdi} categories.\label{tab2}}

\begin{tabular}{|p{50pt}|p{200pt}|}
\hline
\textbf{Category} & \textbf{Annotations} \\
\hline
N & Normal, Left/Right bundle branch block, Atrial escape, Nodal escape \\
\hline
S & Atrial premature, Aberrant atrial premature, Nodal premature, Supra-ventricular premature \\
\hline
V & Premature ventricular contraction, Ventricular escape \\
\hline
F & Fusion of ventricular and normal \\
\hline
Q & Paced, Fusion of paced and normal, Unclassifiable \\
\hline
\end{tabular}
\end{table}

\begin{table}
\centering
\caption{Summary of the preprocessed MIT BIH and PTB dataset.\label{tab1}}
\begin{tabular}{|p{40pt}|p{60pt}|p{60pt}|p{60pt}|}
\hline
\textbf{Dataset} & \textbf{No of classes} & \textbf{No of samples} & \textbf{No of features} \\
\hline
PTB & 2 & 14552 & 187 \\
\hline
MIT-BIH & 5 & 109446 & 187 \\
\hline
\end{tabular}
\end{table}

\subsubsection{Data Exploration and Imbalance Process}
\begin{figure*}[!htb]%
    \centering    \subfloat{{\includegraphics[width=0.47\textwidth]{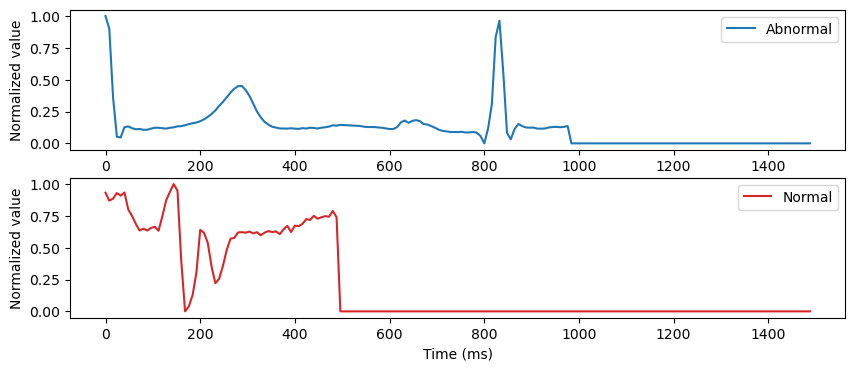} }}%
    \qquad
\subfloat{{\includegraphics[width=0.47\textwidth]{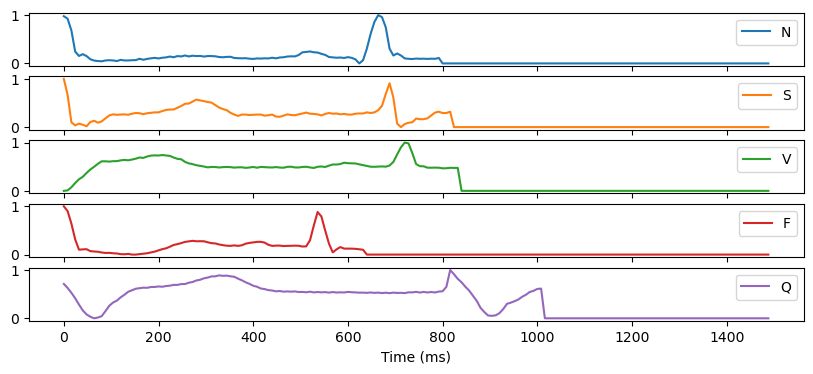} }}%
    \caption{Heartbeat signals from each class of the PTB and MIT-BIH ECG datasets.\label{fig11}}
    \label{fig:example2}%
\end{figure*}

\noindent The variations in amplitude at different times, as shown in Figure \ref{fig11}, highlight the unique characteristics  of each class within the PTB and MIT-BIH datasets. The plot shows that the signals eventually drop to zero, a consequence of the preprocessing applied to these datasets. 
Specifically, the data were cropped, downsampled, and, when necessary, padded with zeroes to achieve a uniform dimension of 187. For more detailed information on this preprocessing, see \cite{preprocess}.

\begin{figure*}[!htb]
    \centering    
\subfloat{{\includegraphics[width=0.35\textwidth]{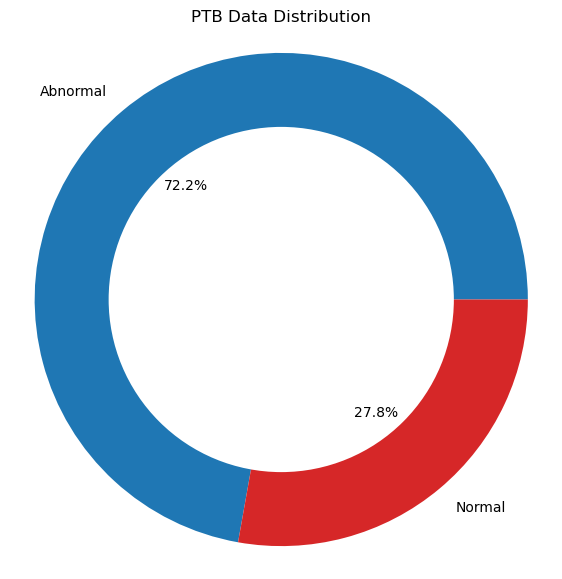} }}%
    \qquad
\subfloat{{\includegraphics[width=0.35\textwidth]{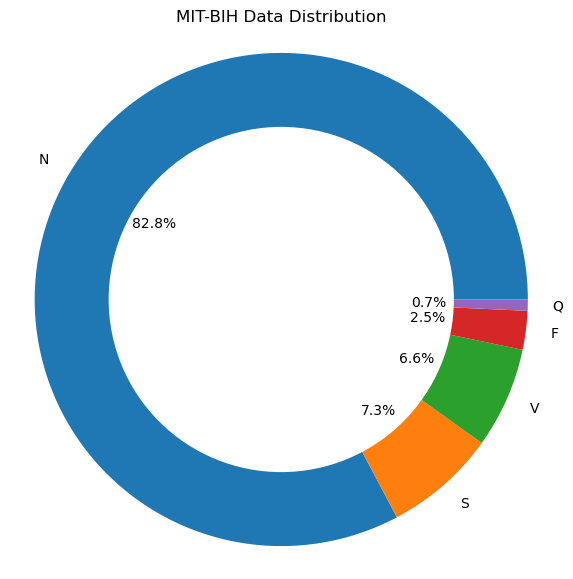} }}%
    \caption{Distribution of heartbeats in different classes for the PTB and MIT-BIH ECG datasets.\label{fig22}}
\end{figure*}

\noindent The distribution of the heartbeats across different classes for the PTB and MIT-BIH datasets is not even as depicted in Figure \ref{fig22}. In the PTB dataset, the number of heartbeat samples labeled as 'Abnormal' significantly exceeds those labeled as 'Normal.' Similarly, in the MIT-BIH dataset, the samples in the 'N' class outnumber those in the minority classes. This disparity leads to a significant imbalance in both datasets. Such class imbalance often results in misclassification, as the decision-making process becomes biased towards the majority class. To address this issue, we explored Synthetic Minority Over-Sampling Technique, commonly known as SMOTE.

SMOTE is an oversampling technique that generates synthetic samples by first identifying a sample from the minority class and then selecting 'k' of its nearest neighbors based on the minimum Euclidean distance. The technique involves calculating the difference between the chosen sample and these neighbors. It then generates new samples by multiplying these differences with random numbers between 0 and 1 and adding the result to the original sample \cite{SMOTE}. These new samples have different features than those which they are mimicking, but maintain similar relational characteristics to the original ones. SMOTE specifically targets the minority class in imbalanced datasets to mimic and enhance its representation.

\subsection{Methods}
\subsubsection{Gaussian Processes}
Gaussian Processes (GPs) are non-parametric Bayesian models that allow inference and uncertainty quantification for a response variable. Formally, a Gaussian process is a collection of random variables, any finite subset of which has a joint Gaussian distribution \cite{gpbook}. 

In practice, GPs are mostly used for regression problems, in which case a Gaussian likelihood function is assumed. As a result, a Gaussian process prior together with a Gaussian likelihood gives rise to a posterior Gaussian process over functions, and everything remains analytically tractable. Applying Gaussian Processes (GPs) to binary classification involves the use of a latent variable to distinguish between classes \cite{gpbook}. For instance, suppose 
a GP is assigned a prior over a discriminative function $f_\theta: \mathbb{R}^P \to \mathbb{R}$, with $P$ representing the feature count and $\theta$ being the hyperparameter. The GP is characterized by its mean function $m(\cdot)$ and a covariance function $k_\theta (\cdot, \cdot)$. Conventionally, the mean function is assumed to be known and so without loss of generality, we assume $m(\cdot)\equiv 0$.
\vskip0.05truecm
Assume $\textbf{z}=(z_1,\cdots, z_n)^\top$ represents the known class labels with $f_\theta(\textbf{x}_i)$ denoting the corresponding value of the latent discriminator at a given input $\textbf{x}_i$. The class labels are determined by the sign of the latent GP as follows:
\begin{equation}
    z_i= \mbox{sign}(f_\theta(\textbf{x}_i)) =\begin{cases}
        +1 \quad f_\theta(\textbf{x}_i)>0\\
        -1 \quad f_\theta(\textbf{x}_i)<0.
    \end{cases} 
\end{equation}

\noindent Suppose that $\textbf{f}\in \mathbb{R}^n$ constitute evaluations of a continuous, surrogate discrimination function $f_\theta$ on $X_{train}=\{\textbf{x}_1, \cdots, \textbf{x}_n\}$. Additionally, suppose $\textbf{y}$ represents the actual observations of $f_\theta$ on $X_{train}$ perturbed by homoscedastic Gaussian noise $\epsilon$. The objective is to predict the response of $f_\theta$, denoted as $\textbf{f}_* \in \mathbb{R}^m$, for unobserved test data $X^*_{test} = \{\textbf{x}^*_1, \cdots, \textbf{x}^*_m\}$. The assumption that $f_\theta$ follows a Gaussian Process with mean zero and covariance function $k_\theta(\cdot, \cdot)$ establishes a Bayesian prior model on $\textbf{f}$, the true evaluations of $f$ on 
$X_{train}$, given by:
\begin{align}
    \dfrac{\textbf{y}}{\sigma} &= \textbf{f} + \epsilon,\\ \nonumber
    \textbf{f} &= [f_\theta(\textbf{x}_1), \cdots, f_\theta(\textbf{x}_n)]^\top \sim \mathcal{N}(\textbf{0},K_{\textbf{ff}}),\\ \nonumber
    \epsilon &\sim \mathcal{N}(0, \eta^2I_n).
\end{align}

Here, $K_{\textbf{ff}}$ represents an $n\times n$ positive definite covariance matrix defined on the training data, where $(i,j)$th element is determined by the covariance function $k_\theta(\textbf{x}_i, \textbf{x}_j)$, and $\eta^2$ is the variance of the unbiased homoscedastic noise. Using the definition of GP regression, we have that the joint distribution of the training and test responses $\textbf{y}$ and $\textbf{f}_*$ as:

\begin{equation}
    \begin{bmatrix}
        \textbf{y}\\
        \textbf{f}_*
    \end{bmatrix}
   \sim  \mathcal{N}\Bigg(0, \sigma^2 \begin{bmatrix}
       K_{\textbf{ff}} + \eta^2I_n & K_{\textbf{f}*}\\
       K_{*\textbf{f}} & K_{**}
   \end{bmatrix} \Bigg),
\end{equation}
where $K_{\textbf{f}*} =K_{*\textbf{f}}^\top$ is the cross-covariance matrix between the training and the testing data, and $K_{**}$ is the covariance matrix of the testing data. Furthermore, the posterior distribution of the test response $\textbf{f}^*$ on $X^*_{test}$ is computed as:
\begin{align}\label{eqn3}
    \textbf{f}_*|X_{train}, X^*_{test},\ \textbf{y} &\sim \mathcal{N}(\Bar{\textbf{f}}^*, \sigma^2C),\\ \nonumber
    \Bar{\textbf{f}}^* &\equiv K_{*\textbf{f}}(K_{\textbf{ff}} + \eta^2I_n)^{-1}\textbf{y},\\ \nonumber
    C &\equiv K_{**} - K_{*\textbf{f}}(K_{\textbf{ff}} + \eta^2I_n)^{-1}K_{\textbf{f}*}
\end{align}

The posterior mean $\Bar{\textbf{f}}^*$ given in equation \ref{eqn3} depends on $\textbf{y}$, and since we only have access to the threshold class indicators $\textbf{z}$, we use similar approach as in \cite{galaxy11} to identify $\textbf{z}$ for $\textbf{y}$ in equation \ref{eqn3}. Also, since the posterior distribution given in equation \ref{eqn3} depends on the choice of the kernel, we select the Mat\'{e}rn kernel, which is a stationary and isotropic kernel, see \cite{galaxy11} for more details.

\subsubsection{MuyGPs}
MuyGPs is an approximate method for training stationary GP hyperparameters using leave-one-out cross-validation restricted to local predictions \cite{muygps}. This methodology is derived from the union of two insights: optimization by way of leave-one-out cross-validation which enables the avoidance of evaluating the likelihood, and restriction to the $k$ nearest neighbors of a prediction location which limits the cost of computing the kriging weights $(K_{*\textbf{f}}(K_{\textbf{ff}} + \eta^2I_n)^{-1})$. Muyskens \textit{et al.} \cite{galaxy11} extended the use of an approximate leave-one-out cross-validation method, originally applied to GP regression in \cite{muygps}, to adapt it for classification problem. 
\vskip0.1truecm
Let $\textbf{x}_{N_i}$ be the set of training observations nearest to $\textbf{x}_i$ and $\textbf{z}_{N_i}$ be their corresponding labels.
A MuyGPs model obtains the $i$th nearest neighbors mean GP prediction by 
\begin{equation}
    f_\theta^{NN} (\textbf{x}_i) = K_{i,N_i}K_{N_i,N_i}^{-1}\textbf{z}_{N_i},
\end{equation}
where $K_{i,N_i}$ denotes the cross-covariance between $\textbf{x}_i$ and $\textbf{x}_{N_i}$, and $K_{N_i,N_i}$ denotes the covariance for the points within $\textbf{x}_{N_i}$. 

MuyGPs estimates $\theta$ using cross-entropy loss, also referred to as log loss over a samples batch $B$, a subset of the training data, by maximizing the negative of the loss function, $Q(\theta)$:
\begin{equation}
    \hat{\theta}=\arg \max_{\theta} -Q_B(\theta),
\end{equation}
where the loss function is:
\begin{align}
    Q_B(\theta) = -\sum_{i\in B}\Bigg\{\dfrac{z_i+1}{2}\log\big[\delta(f_\theta^{NN} (\textbf{x}_i))_0\big] 
    + \Big(1-\dfrac{z_i+1}{2}\Big)\log\big[\delta(f_\theta^{NN} (\textbf{x}_i))_1\big]\Bigg\},
\end{align}
    
and $\delta:\mathbb{R}\longrightarrow[0,1]^2$ is a softmax function defined as
$\delta(a)=\Big(\dfrac{e^{a}}{e^{a}+ e^{-a}},\dfrac{e^{-a}}{e^{a}+ e^{-a}}\Big)$.
The MuyGPs method reduces the computational complexity of optimizing GPs from $O(n^4)$ to $O(bk^3)$, where $b$ is the batch count and $k$ is the number of nearest neighbors.

\subsubsection{K-Nearest Neighbors}
K-Nearest Neighbors (KNN) Classifier assigns labels to samples by examining and comparing the labels of nearby samples. This method is particularly effective when the features involved have relatively similar importance and fall within similar ranges, making it a suitable approach for classification tasks. In terms of optimization, the primary hyperparameter that needs to be adjusted in this method is the number of neighboring samples considered for making the classification decision.

\subsubsection{Decision Tree-Based Ensemble Methods}
A Decision Tree breaks down highly complex decision-making processes by making a series of smaller decisions,  leading to a straightforward and interpretable outcome \cite{DT}. However, this approach can suffer from low accuracy and overfitting. To address these issues, Random Forest Classifiers create multiple small decision trees, each trained on random subsets of features and limited samples chosen with replacement \cite{RF}. A further derivation of this concept is Extra Randomized Trees, which chooses samples without replacement and uses non-optimal smaller decisions to enhance data coverage. While this approach generally exhibits higher bias compared to Random Forests, it benefits from lower variance and reduced computation time, while maintaining comparable accuracy. 
Here, we optimize parameters such as the number of trees in the forest, the minimum number of samples required for a decision node, and the size of the feature subsets. 
\subsubsection{Neural Networks}
In our analysis, we also considered a Convolutional Neural Network (CNN) as they are widely used in ECG signal analysis. The CNN model was designed as a 1D-CNN comprising two convolutional layers, each succeeded by a Max-pooling layer. Following these, the model integrates three fully connected layers, all employing rectified linear unit (RELU) activation functions, to enhance non-linear learning capabilities. For optimization, we selected cross-entropy as the loss function. While previous studies, such as \cite{ecg_cnn1} and \cite{ecg_cnn2}, have utilized deeper CNN architectures for classifying ECG signals, our model adopts a comparatively simpler architecture (see Appendix). This decision was made to achieve a balance between accuracy and computational efficiency.
\vskip0.05truecm
In addition to the CNN model, we also explored a hybrid approach that combines the CNN with MuyGPs. This model first utilizes a CNN to extract high-level spatial features from ECG signals. The extracted features are then fed into MuyGPs. This hybrid model leverages the powerful feature extraction capabilities of CNNs and also approximates a Gaussian distribution over the embeddings, enabling the estimation of uncertainty in the predictions. The detailed architecture of our models, and the hyperparameters used can be found in the Appendix.
\subsection{Tuning the Machine Learning Algorithms}
To optimize hyperparameters for Extra Trees and Random Forest,  we used Marine Predator Algorithm (MPA), \cite{marine}. MPA mimics natural predator behavior in its optimization strategy, which is divided into three distinct phases.
In the first and second phases, the algorithm uses Lévy flights \cite{levy} to explore the solution space, thereby avoiding local minima and maxima. In the subsequent second and third phases, it uses Brownian movements to find the optimal peaks of the solutions identified in the previous phase(s) \cite{marine}. For the tree-based models (random forest and extra trees), we optimized the number of trees, the minimum samples to leaf, and the number of features for each tree. Detailed information about the optimized hyperparameters and model architectures can be found in the Appendix.
\section{Results}\label{sec4}
In this section, we present results on the performance of several models considered for this paper on the PTB (binary-class) and MIT-BIH (multi-class) datasets.

\begin{table*}
\centering
\caption{Accuracy(\%) of different models on PTB.\label{perf1_ptb}}
\begin{tabular}{|p{36pt}|p{38pt}|p{24pt}|p{34pt}|p{52pt}|p{24pt}|p{34pt}|p{34pt}|p{24pt}|}
\hline
Model   & MuyGPS  &  CNN & Extra Trees & Logistic Regression & GP & CNN-Muy & RForest  & KNN \\
\hline
Raw  & 97.84     &\textbf{98.39}    &  97.63    &  83.17       &  91.34   & \textbf{99.18 }  & 96.53   & 93.95\\
\hline
SMOTE &  97.84    &  \textbf{98.35}  &    97.32       &  79.25       & 91.79    & \textbf{98.59}   & 96.63   & 93.20  \\
\hline
Trunc &   \textbf{98.52}  & \textbf{98.35}  &  97.63   & 82.38       & 94.23    &\textbf{98.94}   & 96.56 & 97.18\\
\hline
TruSM & \textbf{98.18}     & \textbf{98.35}  &  97.70    & 78.01   & 94.37    & \textbf{98.94}   & 96.36  & 96.46\\
\hline
\end{tabular}
\newline
\noindent{\footnotesize{\textsuperscript{1} Raw represents the raw data, TruSM represents a combination of of SMOTE and truncation, Trunc is the truncated data.}}
\end{table*}

\begin{figure*}[!htb]
    \centering    
    \subfloat
    {{\includegraphics[width=0.47\textwidth]{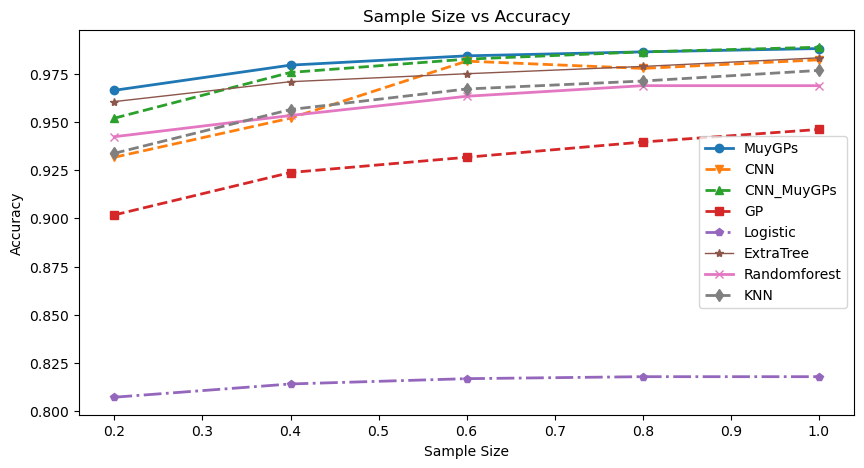} }}%
    \qquad
    \subfloat
    {{\includegraphics[width=0.47\textwidth]{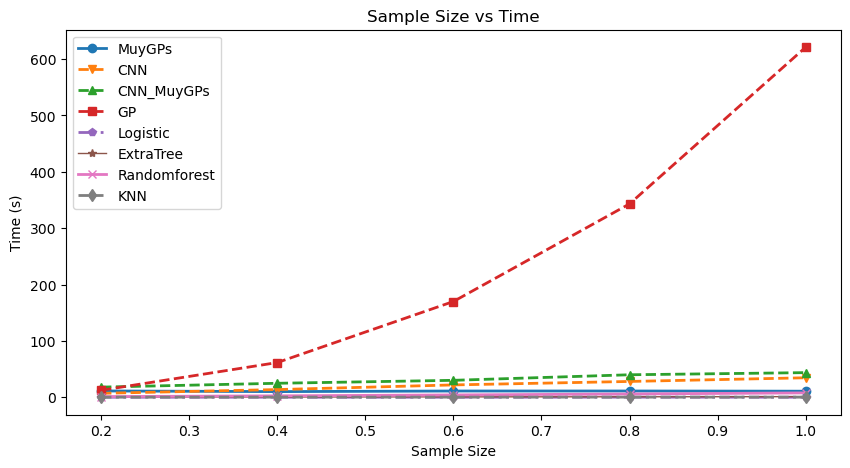} }}%
    \caption{Accuracy vs. Sample Size (left) and Time vs. Sample Size (right) for Different Models on the truncated PTB dataset.}%
    \label{ptb_comp}%
\end{figure*}

\begin{figure*}[!htb]%
    \centering    
    \subfloat
    {{\includegraphics[width=0.47\textwidth]{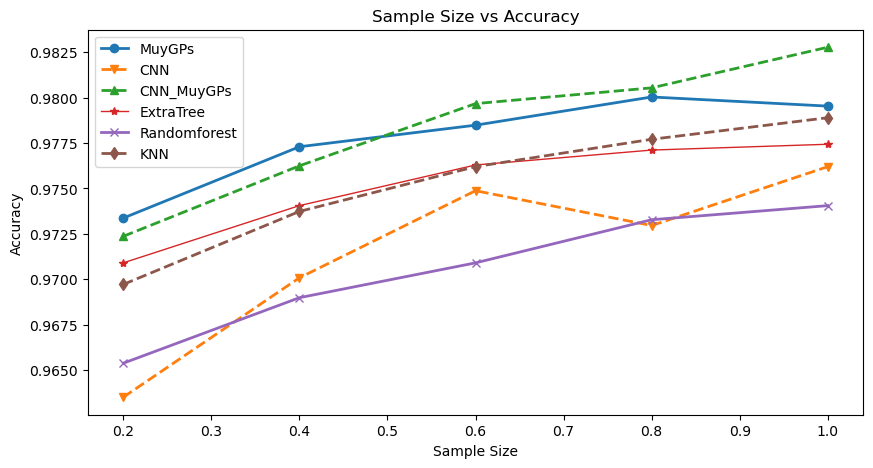} }}%
    \qquad
    \subfloat
    {{\includegraphics[width=0.47\textwidth]{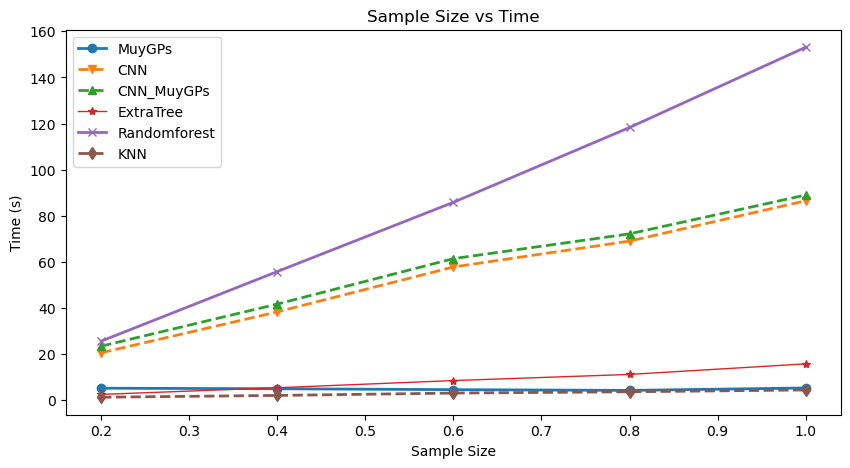} }}%
    \caption{Accuracy vs. Sample Size (left) and Time vs. Sample Size (right) for different models on the truncated MIT-BIH dataset.}%
    \label{mit_comp}%
\end{figure*}

\begin{table}
 \centering
\caption{Accuracy(\%) of different models on MIT-BIH \label{perf1_mit}}
    \begin{tabular}{|c|c|c|c|c|c|c|}
    \hline
     Model   & MuyGPS  &  CNN & Extra Trees  & CNN-Muy & RForest  & KNN \\
     \hline 
      Raw  &   97.98 &  \textbf{98.22} &97.92 & \textbf{98.33}& 97.62   & 97.64 \\
    
      SMOTE &   97.95 & 97.86 & \textbf{98.05} & \textbf{98.34} & 97.70  & 97.30\\
      Trunc &   \textbf{98.09} & 97.74 & 97.74 & \textbf{98.16} & 97.41  & 97.89\\
      TruSM &   97.97 & 97.77 & 97.88 & \textbf{98.29} & 97.57  & 97.50\\
     \hline
    \end{tabular}
\newline
\noindent{\footnotesize{\textsuperscript{1} Raw represents the raw data, TruSM represents a combination of of SMOTE and truncation, Trunc is the truncated data.}}
\end{table}
\vskip0.05truecm
\noindent Tables \ref{perf1_ptb} and \ref{perf1_mit} summarize the accuracies of the different models considered in this study on the
various versions of the datasets ("Raw", "SMOTE", "Trunc" and "TruSM"). The "Raw" represents the entire dataset. "SMOTE" denotes the full dataset with the SMOTE imbalance correction applied, using five nearest neighbors to generate synthetic data points. The proportion of synthetic data points added to the minority class relative to the majority class was 80\% for the PTB dataset and 8\% for the MIT-BIH dataset. These settings were optimized using best validation set performance. "Trunc" represents truncated data which comprises of the first $80$ features, and "TruSM"  represents the truncated data with SMOTE imbalance technique being applied. 
\vskip0.05truecm
Best,
From the tables, we observe that truncating the datasets improved the performance in most of the models considered. This may be as a result of the models focusing on the most relevant segments of the data during the training, and also because truncation eliminates padding around the end of the signals, reducing potential noise and distortions. On the contrary, applying SMOTE both on the raw and truncated PTB dataset has less or no effect on the performance of the models. Although applying smote resulted in a slight performance increase for some models on the  MIT-BIH dataset, it should be noted that there is a computational time cost trade-off when incorporating synthetic data into an already large dataset. Therefore, we choose the truncated datasets for the rest of the analysis we will be performing.
\vskip0.05truecm
\noindent Figures \ref{ptb_comp} and \ref{mit_comp} compare the accuracies and training/testing time of models across various sizes of training sets derived from the PTB and MIT-BIH datasets. We obtained these training set sizes by sampling from the training sets with ratios $\{0.2, 0.4, 0.6,0.8, 1\}$. Subsequently, the classifiers were trained on these subsets and were then evaluated on a fixed held out test sets. The hardware specifications used to run the comparisons is listed in the Appendix. From the results, we observe that MuyGPs compares favorably with other models in terms of accuracy while significantly reducing the computational time complexity of the traditional Gaussian process model. Also, its training/testing time is comparable to the other computationally efficient models considered in this study. Notably, the extensive computational requirements of Gaussian Process model made its application impractical on the MIT-BIH dataset, leading to its exclusion from this particular analysis.
Furthermore, these figures highlight that MuyGPs has the best performance on small training sizes, indicating its effectiveness in scenarios where data availability is limited.
\vskip0.05truecm
In addition, the CNN-MuyGPs model, which integrates CNN embeddings with the MuyGPs framework, enhances the accuracy of both the CNN and MuyGPs, particularly for larger training sizes. This improvement in accuracy is attributed to the model leveraging the advantages of both models: CNN serves as a preprocessing tool to extract meaningful features, while MuyGPS is utilized to analyze the relationships within these features. However, combining these models leads to an increase in computational cost which can be seen from the figures.

\section{Uncertainty Estimation}\label{sec3}
In this section, we discuss the uncertainty estimation for the different models examined in our study. In order to focus on the most important models we limit our discussion to our top five models: MuyGPs, CNN-MuyGPs, CNN, ExtraTree, and Gradient Boost.

\subsection{MuyGPs}

In Gaussian processes, the estimation of uncertainty in predictions is traditionally achieved through conditional distributions. However, the application of this concept to classification tasks, particularly in leveraging latent information for uncertainty estimation, presented challenges. The MuyGPs model addresses this by incorporating latent uncertainties within a hypothesis testing framework \cite{galaxy11}. It introduces a third class labeled "ambiguous" for instances where the latent Gaussian process is non-significantly different than the class decision boundary (set at $0$). This approach allows for the exclusion of ambiguous heartbeat signals, enhancing the classifier's accuracy. Moreover, it flags these uncertain cases for potential manual inspection or further analysis.
\vskip0.1truecm
\noindent Suppose for a binary classification data, we assign $\textbf{z}_i = +1$ to denote a normal observation and $\textbf{z}_i = -1$ to indicate an abnormal one. Then to evaluate the test data, the latent predictions less than $0$ are classified as abnormal, while those greater than $0$ are considered normal. Consider the hypotheses
$  H_0 : f_i = 0$ versus $H_1:f_i\
ne 0 $. In the testing framework, a Type I error ($\alpha$) is when $0$ is not contained in a prediction interval, but the classification is incorrect. On the other hand, a Type II error ($1 -\beta$) occurs when $0$ is within the prediction interval, but the prediction is in the correct classification. To estimate the prediction variance, we minimize the function, $\alpha + 1 -\beta$, in these tests, see \cite{galaxy11} for several example functions that yield a range of possible values.

\subsection{Neural Network}
One of the methods commonly used for uncertainty estimation in neural networks is Monte Carlo Dropout simulation (MCDS). Traditionally, dropout is used as a model regularization technique to avoid overfitting. It works by randomly removing neurons with a given probability p, during training. However, in MCDS dropout is turned on during inference to produce slightly different outputs for a given test point. This approach allows our CNN model, which integrates dropout before each layer, to produce a variety of outputs for the same test point. During testing, we keep dropout on and perform $n=30$ runs resulting in 30 slightly varied predictions per test point. The uncertainty in test prediction is then studied using the mean and variance of these predictions.
\subsection{Other Machine Learning Models}
None of the remaining models have methods to derive probabilities, but they do have ways to estimate confidence. Extra Trees and Random Forest take the mean probability of each tree in their forest, with those probabilities being the fraction of samples in the same class on that leaf \cite{RF}. We run each model $n=30$ times with different randomness settings, which gives us 30 models with slight differences to calculate the mean and variance for our probability intervals.
 
\vskip0.05truecm
\noindent Following the uncertainty estimation methods discussed above, we estimate the prediction variance for each test point in each signal. For each model, we use these predictions and their associated variances to construct  confidence intervals for each test point. Precisely, for different values of  $\tau$, $\tau := [0.994, 1.28, 1.64, 1.96, 2.58]$, we estimate the prediction intervals
\begin{equation}
   \mathcal{I}=\hat{f}(x_i)\pm \tau \hat{\sigma}. 
\end{equation}
These $\tau$ values correspond to $68\%$, $80\%$, $90\%$, $95\%$ and $99\%$ confidence levels of the prediction interval, respectively. We classify intervals that  contain $0$ (for MuyGPs) or $0.5$ (for other models)  as ambiguous, otherwise we use the classification given by each classifier.

\begin{figure*}[t]%
    \centering    
    \subfloat
    {{\includegraphics[width=0.47\textwidth]{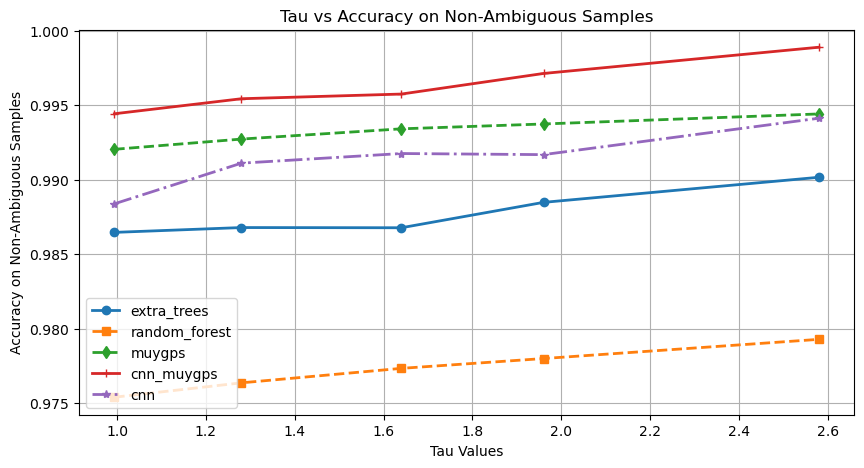} }}%
    \qquad
    \subfloat
    {{\includegraphics[width=0.47\textwidth]{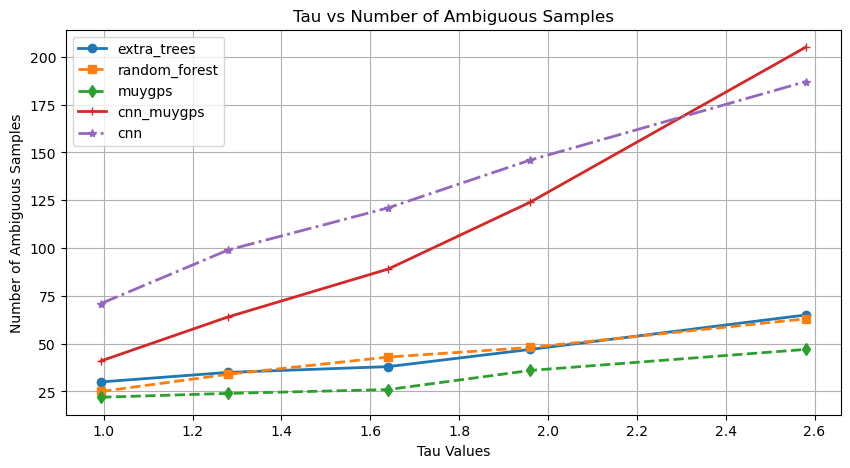} }}%
    \caption{Tau Vs. Accuracy (left) and Tau Vs. Number of Ambiguous signals (right) on PTB-dataset. The tau values  $[0.994, 1.28, 1.64, 1.96, 2.58]$ correspond to $68\%$, $80\%$, $90\%$, $95\%$ and $99\%$ confidence levels of the prediction interval respectively.}%
    \label{fig1}%
\end{figure*}

\begin{figure*}
    \centering    
    \subfloat
    {{\includegraphics[width=0.47\textwidth]{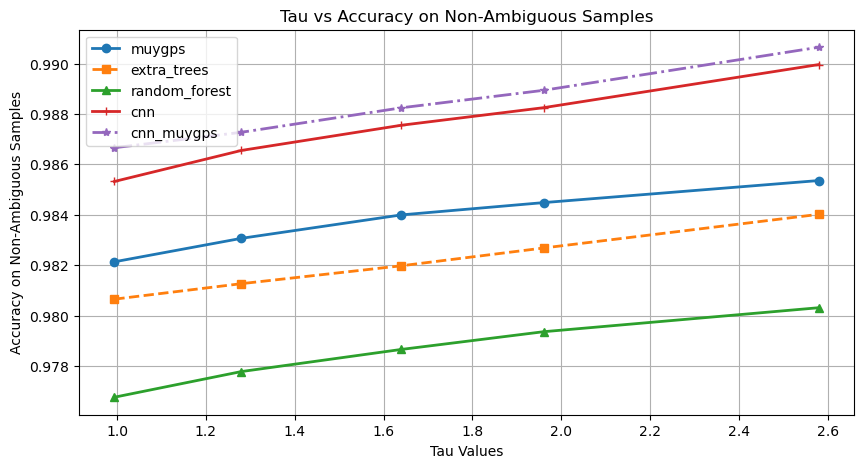} }}%
    \qquad
    \subfloat
    {{\includegraphics[width=0.47\textwidth]{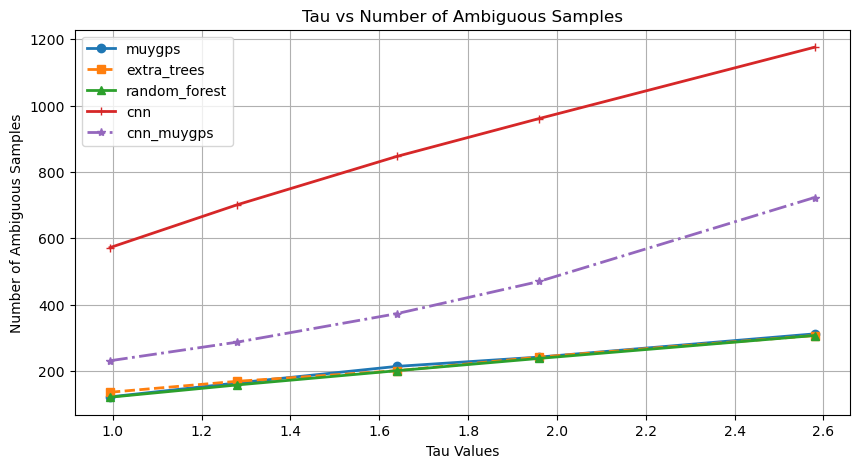} }}%
    \caption{Tau Vs. Accuracy (left) and Tau Vs. Number of Ambiguous signals (right) on MIT-BIH dataset.}%
    \label{fig2}%
\end{figure*}

\begin{figure*}
    \centering    
    \subfloat
    {{\includegraphics[width=0.47\textwidth]{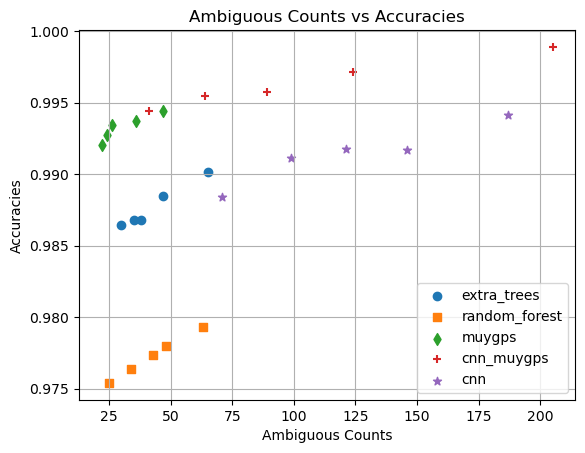} }}%
    \qquad
    \subfloat
    {{\includegraphics[width=0.47\textwidth]{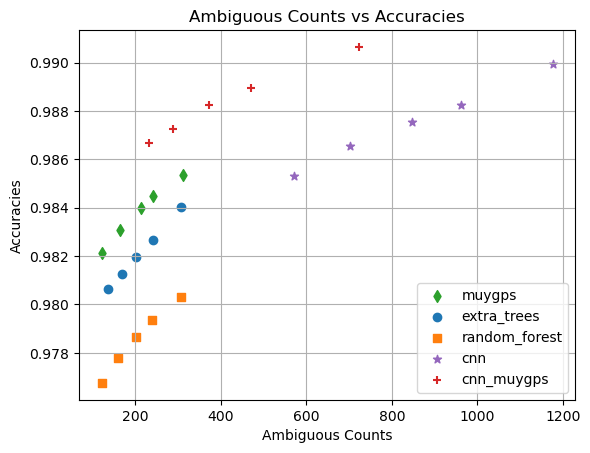} }}%
    \caption{Number of ambiguous signals vs. Accuracy for PTB Dataset (left) and MIT-BIH dataset (right)}%
    \label{scata}%
\end{figure*}

Figure \ref{fig1} presents a comparison of model accuracies of the non-ambiguous samples and the number of ambiguous signals across various $\tau$ values on the PTB dataset.
The plot shows that MuyGPs not only achieves high accuracy on the non ambiguous signals but also eliminates the smallest quantity of ambiguous signals. This outcome is highly desirable as it ensures confident predictions with minimal necessity for further verification by human operators. Meanwhile, CNN-MuyGPs attains the best accuracy on the non-ambiguous signals, however, it assigns a high number of ambiguous signals, thereby increasing the verification burden on human operators. It is important to note that the computational time required by MuyGPs and CNN-MuyGPs for uncertainty quantification is significantly less compared to other models. This is due to the fact that obtaining the variance for other models requires a simulation, while it is inherent for MuyGPs.
\vskip0.05truecm
The approach we utilized for discerning ambiguous signals in the binary dataset is not directly transferrable to multi-class datasets. For a multi-class data, we adopted a one-versus-all (OVA) \cite{gal} approach to binarize the multi-class into binary classes. Then we perform the uncertainty estimation of the sub-binary classes obtained following the procedures as in the binary data. That is, for a given $\tau$ value, we identify the ambiguous signals for each sub-binary class. Then we remove the collective set of these signals and calculate the accuracies for the non-ambiguous signals. Concretely, given a dataset $D$ with classes \(\{C_1, C_2, \ldots, C_n\}\), for each class \(C_i\), we deploy an OVA classifier \(\hat{f}_i\), where for example \(\hat{f}_i(x)\) ranges from \(-1\) to \(1\) (as in MuyGPs). For a given $\tau$ value and a class \(C_i\), if $0 \in \mathcal{I}_i=\hat{f}_i(x)\pm \tau \hat{\sigma} \text{, then } x \text{ is uncertain for } C_i$. Then the total uncertain classification is determined by the union of the uncertain classifications from each class.
\vskip0.05truecm
Figure \ref{fig2} demonstrates the performance of the models under consideration on the MIT-BIH dataset. From the figure, it is evident that removing the ambiguous signals improves the accuracy of all models considered. Similar to the binary case, the accuracy of MuyGPs increases while removing a relatively small number of ambiguous signals. Both CNN and CNN-MuyGPs achieve high accuracy on the non-ambiguous signals. However, CNN-MuyGPs has a better performance than CNN while reducing the number of ambiguous signals as well as computational time,  which are desirable for practical applications. 
Figure \ref{scata}, summarizes the trade-off between number of ambiguous signals and accuracy on the non-ambiguous signals for both PTB and MIT-BIH datasets. It is optimal to be in the top-left corner of these plots, and we can see that MuyGPs and CNN-MuyGPs are the best performers.

\section{Conclusion}\label{sec5}
In this study, we explored an approximate Gaussian process, MuyGPs, on classifying binary and multi-class ECG datasets. MuyGPs outperformed the state-of-the-art machine learning methods such as CNN, Randomforest, Extra trees, etc, while requiring fewer training samples and significantly reducing the time complexity compared to the traditional GP classifier. 
In addition, we trained MuyGPs using features extracted from a CNN model, CNN-MuyGPS, and it was shown CNN-MuyGPs improve MuyGPs in accuracy with trade-offs in time complexity and larger training samples. 

Furthermore, MuyGPs also performed superior performance in estimating uncertain ECG signals classifications. Particularly, It successfully removed minimal ambiguous signals while maintaining a high accuracy compared to other models considered in the study.

\vskip0.4truecm

This work was performed under the auspices of the U.S. Department of Energy by Lawrence Livermore National Laboratory under Contract DE-AC52-07NA27344 with IM release number LLNL-JRNL-860415 and was supported by the LLNL-LDRD Program under Project No. 22-ERD-028.

\begin{appendix}

\section{Appendix}

The analysis of the models studied in this paper were implemented on Python on a Linux operating system with CPU processor-AMD, Ryzen 5 5600 6-Core and 32GiB RAM. The MPA algorithm was used to obtain optimal hyperparameters for Extra tress and Random forest models. 

\noindent The GP model used in this paper was implemented using the sklearn library and we did not use the noise model due to the time complexity. Tables \ref{appen2} and \ref{appen1} summarize the specifications of the models used in this study. 

\begin{table}
\caption{ Model Specifications \label{appen2}}
    \centering
    \begin{tabular}{|p{50pt}|p{250pt}|}
    \hline
     Model   &  Training Configuration \\
     \hline
          MuyGPs & \begin{itemize}
         \item Mat\'ern ( nu=1.5, l=1, noise=$10^{-5}$)
         \item Nearest Neighbor = PTB-50, MIT-35
     \end{itemize} \\
     \hline
     CNN &  \begin{itemize}
         \item 2 Conv layers (64 and 68 channels) with 3$\times $3 kernel,  each followed by  Max Pool (2$\times$ 2) 
         \item 3 Fully Connected Layers with widths of 128, 80, and 'number of classes' neurons  
         \item Cross Entropy loss and Relu activation 
         \item Epochs - 30 (ptb), 10 (mit) 
         \item Learning Rate - 0.001
         \item Optimization - Adam
     \end{itemize}
     All layers are followed by Dropout (p=0.1) \\
     \hline
     CNN-MuyGPs & \begin{itemize}
         \item 80 Feature embeddings from 2nd Fully Connected layer of CNN inputted to MuyGPs model
     \end{itemize}\\
     \hline
    \end{tabular}
    \end{table}
    
\begin{table}
\caption{ Model Specifications continued \label{appen1}}
    \centering
    \begin{tabular}{|p{50pt}|p{250pt}|}
    \hline
     Model   &  Training Configuration \\
      \hline
     Log Reg & \begin{itemize}
         \item Loss (L2)
         \item Iteration =100
     \end{itemize} \\
     \hline
     Extra Tree & \begin{itemize}
     \item Estimators - 159 
     \item Num Features -  172
     \item Splitting - Gini Impurity
     \item Leaf Samples Min - 1
     \item Max Depth - None
     \end{itemize},  \\
     \hline
     Random Forest & \begin{itemize}
     \item Estimators - 360 
     \item Number of Features -  1
     \item Splitting - Gini Impurity 
     \item Leaf Samples Min - 1
     \item Max Depth - None
     \end{itemize} \\
              \hline
     GP   &   \begin{itemize}
         \item Mat\'ern (nu=1.5, l=1) 
     \end{itemize}\\
     \hline
          KNN & \begin{itemize}
         \item 3 NN 
     \end{itemize}\\
     \hline
     \end{tabular}
     \end{table}

\end{appendix}

\bibliographystyle{unsrtnat}
\bibliography{Our_Manuscript}  






\end{document}